\documentclass[11pt,a4paper]{article}

\usepackage[utf8]{inputenc}
\usepackage[T1]{fontenc}
\usepackage{amsmath,amssymb}
\usepackage{graphicx}
\usepackage[margin=1in]{geometry}
\usepackage{authblk}
\usepackage{abstract}
\usepackage[round]{natbib}
\usepackage{hyperref}
\usepackage{booktabs}
\usepackage{caption}
\usepackage{subcaption}
\usepackage{float}
\usepackage{xcolor}

\usepackage{setspace}

\hypersetup{
	colorlinks=true,
	linkcolor=blue,
	citecolor=blue,
	urlcolor=blue
}

\title{\bfseries Solar-Cycle Variation of Newly Emerging Coronal Holes during Solar Cycle 24}

\author[1]{Akash Vinod Shirke}
\author[2]{Khagendra Katuwal}
\affil[1]{Department of Physics, UPES, Energy Acres, Dehradun, Uttarakhand, 248007, India}
\affil[2]{Department of Astronomy, New Mexico State University, Las Cruces, NM 88003, USA}

\date{}

\begin{document}
	
	\maketitle
	
	\begin{abstract}
		\noindent
	Coronal holes (CHs) are relatively dark regions of the solar corona characterized by predominantly open magnetic field lines. They are considered to be the source of high-speed solar wind streams. We report a long-term (2010–2019), manually compiled catalogue of newly emerging coronal holes (CHs), identified from daily Coronal Hole Identification via Multi-thermal Emission Recognition Algorithm (CHIMERA) maps provided through SolarMonitor, covering Solar Cycle 24. In contrast to the existing area-based datasets, our catalogue archives the daily emergence rate of new CHs, categorized based on latitudinal zones: northern hemisphere ($+20^{\circ} < \mathrm{lat} \leq +90^{\circ}$), southern hemisphere ($-90^{\circ} \leq \mathrm{lat} < -20^{\circ}$), and equatorial zone ($|\mathrm{lat}| \leq 20^{\circ}$). We find that the emergence rate of equatorial CHs is anti-correlated with the International Sunspot Number (SSN) at monthly cadence (Pearson $r = -0.523$, $p = 6.2 \times 10^{-9}$), suggesting that the formation of lower-latitude open magnetic flux is suppressed near solar maximum. The North-South asymmetry index (N-S AI) of newly emerging CHs is on average south-dominated over the full cycle. Still, it reverses to north-dominated during 2013-2015, coincident with the second peak of the Solar Cycle~24 double maximum. Notably, the hemispheric asymmetry of CHs is also anti-correlated with the hemispheric asymmetry of SSN (Pearson $r = -0.461$, $p = 5.2 \times 10^{-7}$), a previously unreported result suggesting that the emergence of CHs preferentially favours the hemisphere opposite to that of greater sunspot activity. The total emergence rate of CHs is suppressed at solar maxima ($1.190 \pm 0.039$ CH day$^{-1}$) and highest during the declining phase ($1.420 \pm 0.040$ CH day$^{-1}$). These findings offer new observational constraints on the relationship between the formation of coronal holes (CH), the evolution of hemispheric open flux, and the solar dynamo mechanism.
	\end{abstract}
	

	\section{Introduction}
	\label{sec:intro}
	When we observe the solar corona in extreme-ultraviolet (EUV) and soft X-ray bands, we see dark regions with lower density and temperature than the surrounding quiet corona. They are associated with open magnetic field lines, allowing plasma to escape into the heliosphere \citep{cranmer2009}. The CHs are regions of predominantly unipolar magnetic flux with open magnetic field lines in the solar corona \citep{1982SoPh...79..149H,2026ApJ...999...63K}. CHs are also the primary drivers of the fast-originating solar wind ($>400$ km s$^{-1}$) \citep{bale2023} and of recurrent geomagnetic activity governed by corotating interaction regions (CIRs) \citep{tsurutani2006}. Understanding the formation, evolution, and distribution of CHs over the heliographic latitudes and hemisphere is subsequently important both for solar dynamo theory and for space-weather forecasting.\\
   During solar minimum, large and persistent CHs predominantly occupy the two polar regions, where the solar magnetic field is largely open and dominated by a single magnetic polarity. As the solar cycle progresses toward maximum, the polar CHs shrink and may eventually disappear as the polar magnetic fields undergo reversal, while CHs increasingly occur at lower and middle heliographic latitudes. Many of these non-polar CHs are associated with the dispersal and decay of active-region magnetic flux  \citep{bravo1997}.
   Multiple research studies have utilized automated detection algorithms to track and keep a record of the area and number of CHs throughout one or more solar cycles \citep{garton2018, verbeeck2014, 
heinemann2019, harvey2002}, primarily finding (i) an anti-correlation between the sunspot number (SSN) and coronal hole area at lower latitudes, since active regions both compete for, and gradually supply, open flux to the equatorial belt \citep{lowder2017}; and (ii) a sustained North-South (N-S) asymmetry in coronal hole area, which require not to track the N-S asymmetry of sunspots \citep{nakagawa2019,mcintosh2014database}.\\
    Specifically, \citet{nakagawa2019} observed that the lower-latitude CH area in the southern hemisphere surpassed that in the northern hemisphere during both Solar Cycles 23 and 24, whereas \citet{andreeva2021} reported a similar persistent N-S asymmetry limited to the polar region CH areas. The McIntosh Archive analysis of \citet{mcintosh2014database} observed that the asymmetry index of CH area varies between positive and negative values with no long-duration consistent hemispheric predominance and is almost uncorrelated with the asymmetry of sunspot area - a result that, as we discuss in Section~\ref{sec:asymmetry}, our emergence-count analysis both indicates and fine-tunes temporally.\\
    Nearly all of the research studies are based on the number or the area of CHs present on the solar disk at a given time (a daily recorded image), obtained from the automated segmentation pipelines such as CHIMERA \citep{garton2018}, SPoCA \citep{verbeeck2014}, CATCH \citep{heinemann2019}, or the McIntosh Archie’s synoptic maps data \citep{mcintosh2014database, hewins2020}. The rate at which new coronal holes (CHs) emerge on a given day is a related but unique quantity: it is less sensitive to the lifespan and growth/decay of the area of a given coronal hole (CH) and more explicitly linked to the rate of new open-flux formation.\\
    At present, a manually validated, decade-long catalogue of newly emerging coronal hole (CH) events, classification based on different hemispheric regions (discussed in ~\ref{sec:data:ch}), has not previously been published for Solar Cycle 24. In this research work, we present such a catalogue, developed by manual inspection of the daily CHIMERA-derived coronal hole (CH) maps provided through the SolarMointor service \citep{garton2018, solarmonitor}, covering 2010~October~5 to 2019~December~31 – primary the full observable range of Cycle ~24 (minimum period: December~2008; maximum period: April~2014, smoothed SSN $\approx 116$; end: December~2019; \citealp{silso2024}).
    \bigskip
    \\
    We use this catalogue to investigate three main questions that, to our knowledge, have not previously been examined using an emergence-rate catalogue of this kind. First, we examine how the daily and monthly emergence rates of equatorial coronal holes (CHs) are related to the International Sunspot Number (SSN). Second, we compare the north--south (N--S) asymmetry of newly emerging CHs with the corresponding N--S asymmetry of the SSN and quantify the relationship between them. Finally, we investigate whether the emergence rate of CHs, both globally and within individual hemispheric zones, varies systematically across the different phases of Solar Cycle 24.

  The paper is structured as follows: Section~\ref{sec:data} describes the CHIMERA-based CH catalogue, the SSN dataset used for comparison and correlation, and our treatment of missing observations. Section~\ref{sec:methods} introduces the statistical measures used throughout the paper, including the asymmetry index and the solar-cycle phase boundaries. Section~\ref{sec:results} presents the results addressing each of the three questions outlined above. Section~\ref{sec:discussion} discusses these findings in the context of previous area-based studies, and Section~\ref{sec:conclusion} summarizes our main conclusions.    
	\section{Data}
	\label{sec:data}
	
	\subsection{Coronal hole catalogue}
	\label{sec:data:ch}
Our dataset is a manually compiled catalogue of newly emerging daily coronal-hole counts covering the period from 2010 October 5 to 2019 December 31, spanning most of Solar Cycle~24. For each day, we visually inspected the coronal-hole segmentation maps generated by the Coronal Hole Identification using Multi-thermal Emission Recognition Algorithm \citep[CHIMERA;][]{garton2018}, as distributed through the SolarMonitor archive.\footnote{\url{https://www.solarmonitor.org}} The CHIMERA algorithm identifies coronal-hole (CH) boundaries using intensity ratios among three extreme-ultraviolet (EUV) passbands of the Atmospheric Imaging Assembly (AIA; \citealt{lemen2012}) aboard the Solar Dynamics Observatory (SDO; \citealt{pesnell2012}), centered at 171, 193, and 211~\AA{}. These observations are combined with co-temporal magnetograms from the Helioseismic and Magnetic Imager (HMI) aboard SDO. The HMI magnetograms are used to apply a magnetic-unipolarity criterion \citep{2026ApJ...999...63K} and to exclude filament channels, which may otherwise be misidentified as CHs based solely on their low EUV intensities. CHIMERA has been benchmarked against five other CH-detection frameworks in the community comparison presented by \citet{heinemann2024} and is one of the commonly used automated CH-identification algorithms in the recent literature \citep[e.g.,][]{garton2018,rotter2012,heinemann2019}.
	
	For each day with an available CHIMERA observation map, we manually observed and noted the number of newly emerging Coronal Hole regions, categorized into three hemispheric zones:

	\begin{itemize}
		\item \textbf{Northern zone}: CHs located around the  heliographic latitudes $+20^{\circ} < \mathrm{lat} \leq +90^{\circ}$ (of any size, both mid-latitude and polar CHs) appearing in the northern hemisphere.
        
		\item \textbf{Southern zone}: CHs located around the heliographic latitudes $-90^{\circ} \leq \mathrm{lat} < -20^{\circ}$, including both mid-latitude and polar CHs in the souther hemisphere
        
		\item \textbf{Equatorial zone}: CHs appearing within the low-latitude active region equatorial belt, defined as in the classification scheme applied during data collection ( $|\mathrm{lat}|
		\le 20^\circ$, as this is the major solar active zone).
	\end{itemize}
	
	The sum of all three categories is the Total daily count of CHs. We underline that the Northern and Southern categories, as recorded, include both polar and non-polar CHs in each of the hemisphere. They are best read as ``hemispheric'' instead of strictly ``polar'' counts, as multiple studies separate polar and non-polar CHs explicitly\citep[e.g.,][]{lowder2017,andreeva2021}, we return to this point in Section~\ref{sec:discussion}

   \subsection{Missing data}
	
   Our catalogue contains 3{,}407 calendar days. Of these, 383 days ($11.2$\%) have no recorded CH count because no usable CHIMERA observation map was available, primarily due to data gaps or poor imaging conditions. These days do not indicate that zero CHs were observed; rather, they were treated as missing data (\texttt{NaN}) instead of being assigned a value of zero. All averages reported in this paper, including daily and monthly means, were calculated using only days with valid observations.
   \subsection{Sunspot number data}
	\label{sec:data:ssn}

    Sunspots are regions of strong magnetic field that often appear in pairs of opposite magnetic polarity. They are magnetically distinct from both the quiet Sun and coronal holes \citep{2023AAS...24221404K}. Sunspot magnetic-field strengths are typically of the order of kilogauss, whereas the photospheric magnetic fields associated with the quiet Sun and coronal holes are generally much weaker \citep{2023SPD....5420402K}.
	For comparison with solar activity, we use the International Sunspot Number (SSN) reported by the Sunspot Index and Long-term Solar Observations (SILSO) World Data Center at the Royal Observatory of Belgium \citep{silso2024,clette2014}. We adopted both the total daily and monthly SSN series, as well as the corresponding hemispheric daily and monthly series, and restricted all datasets to the temporal range covered by our CH catalogue. We then compared the equatorial CH counts with the SSN, since sunspots predominantly occur at low heliographic latitudes, and also compared the north--south asymmetry indices derived from the two datasets.

	\section{Methodology}
	\label{sec:methods}
	
	\subsection{Asymmetry index}
	\label{sec:methods:ai}
	
	Following the framework used extensively in the N-S asymmetry literature \citep[e.g.,][]{mcintosh2014database,joshi2007,zhukova2023}, we defined a normalized asymmetry index for any paired hemispheric quantity $N$ (North) and $S$ (South) as 
	\begin{equation}
		\mathrm{AI} = \frac{N - S}{N + S},
		\label{eq:ai}
	\end{equation}
	so that $\mathrm{AI} > 0$ suggests northern dominance and
	$\mathrm{AI} < 0$ suggests southern dominance, with $\mathrm{AI}$ bounded in $[-1, 1]$. We calculate $\mathrm{AI}$ individually for (a) the monthly mean CH counts in each of the hemispheres ("CH AI") and (b) the monthly SILSO hemispheric sunspot number ("SSN AI"), and further compare these two explicitly. Equation~\eqref{eq:ai} is left undefined when $N+S=0$; such months did not arise in either series at monthly cadence.

	\subsection{Solar cycle phase boundaries}
	\label{sec:methods:phase}

	We do the partition of Solar Cycle~24 into three different phases based on the SILSO smoothed monthly SSN \citep{silso2024}: a rising phase starting from our catalogue (2010~October) to the end of the double-peaked rise (2012~march), a maximum phase extending across both peaks of the well-known double maximum of Solar Cycle~24 (2012~April-2015~June; first peak SSN$\,\approx 99$ in April 2014), and a declining phase is from 2015~July to the end of our catalogue (2019~December), which incorporates the long, slow decline feature of Solar Cycle~24 \citep{silso2024}. The specific boundaries used in this paper are: Rising phase (2010~October~1 - 2012~March~31), Maximum phase (2012~April~1 - 2015~June~30), and Declining phase (2015~July~1 2019~December~31). The solar cycle phases are not uniquely specified in this literature; we utilize one consistent scheme here and present it explicitly so that our findings can be reproduced with an alternative convention.

	\subsection{Statistical diagnostics}
	\label{sec:methods:stats}
    
    For all the bivariate comparisons in this research study, we calculate both the Pearson correlation coefficient $r$, defined as
\begin{equation}
    r = \frac{\sum_{i=1}^{n}(x_i - \bar{x})(y_i - \bar{y})}
    {\sqrt{\sum_{i=1}^{n}(x_i - \bar{x})^2 \, 
    \sum_{i=1}^{n}(y_i - \bar{y})^2}},
    \label{eq:pearson}
\end{equation}

\noindent where $\bar{x}$ and $\bar{y}$ are the sample means of two variables, and $n$ is the number of data pairs. This correlation coefficient computes the strength of the linear association between the two variables.

We also calculate the Spearman correlation coefficient $\rho$, specified as the Pearson coefficient applied to the rank-transformed variables:

\begin{equation}
    \rho = \frac{\sum_{i=1}^{n}(R_i - \bar{R})(S_i - 
    \bar{S})}{\sqrt{\sum_{i=1}^{n}(R_i - \bar{R})^2 \, 
    \sum_{i=1}^{n}(S_i - \bar{S})^2}},
    \label{eq:spearman}
\end{equation}

\noindent where $R_i$ and $S_i$ are ranks of $x_i$ and $y_i$, also $\bar{R}$, $\bar{S}$ are their mean ranks. This formulation precisely handles the tied ranks, which emerge naturally in our discrete count data. For the outliers and non-normality, the Spearman rank coefficient is more robust than the Pearson coefficient. Both are reported together with their two-sided $p$-values in this paper.
	
	\section{Results}
	\label{sec:results}
	
	\subsection{Overview of the coronal hole emergence catalogue}
	\label{sec:overview}
	
\begin{figure}[H]
		\centering
		\includegraphics[width=0.97\textwidth]{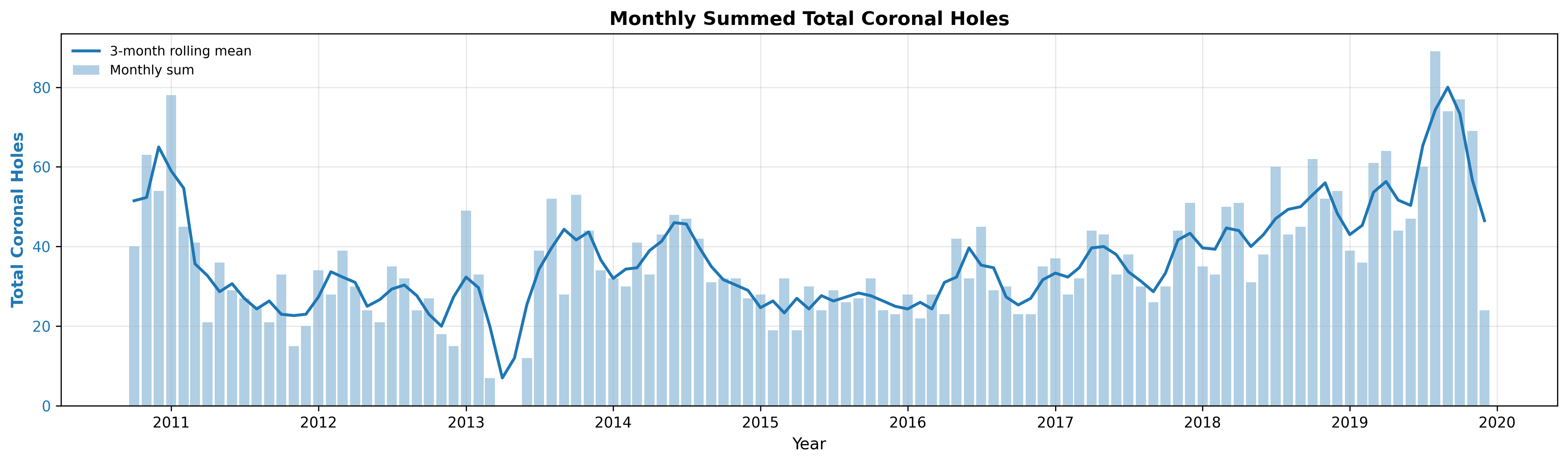}
		\caption{Monthly summed total newly emerging CH count (bars) and three-month rolling mean (blue line), 2010-2019 (Solar Cycle 24). The data show a broad trough in 2013 (solar maximum), with increased counts in the early rising phase (2010-2011) and enhanced trend through the declining phase (2016-2019) toward cycle minimum.}
		\label{fig:total_monthly}
	\end{figure}

\begin{figure}[H]
		\centering
		\includegraphics[width=0.97\textwidth]{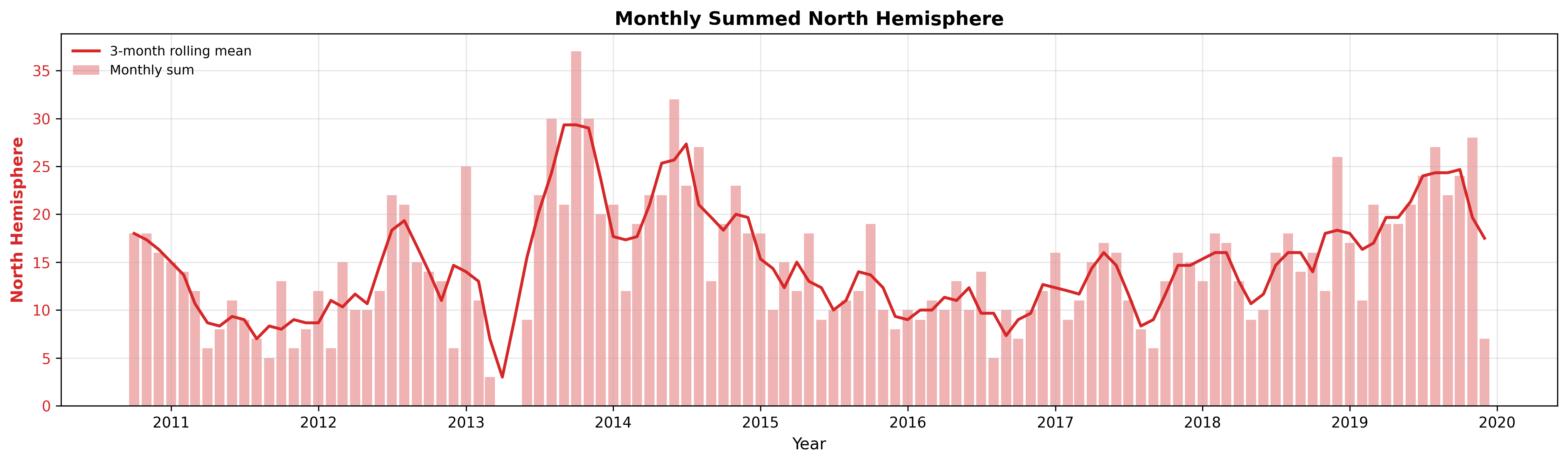}
		\caption{Monthly summed Northern hemispheric newly emerging CH count (bars) and three-month rolling mean (red line). The Northern hemisphere shows a peak during 2013-2014 (monthly sums approaching $\sim$37), corresponding to the period following the Northern hemispheric SSN peak of Cycle 24 (November 2011), before declining gradually through 2015-2016 and recovering from 2018 onward.}
		\label{fig:north_monthly}
	\end{figure}

\begin{figure}[H]
		\centering
		\includegraphics[width=0.97\textwidth]{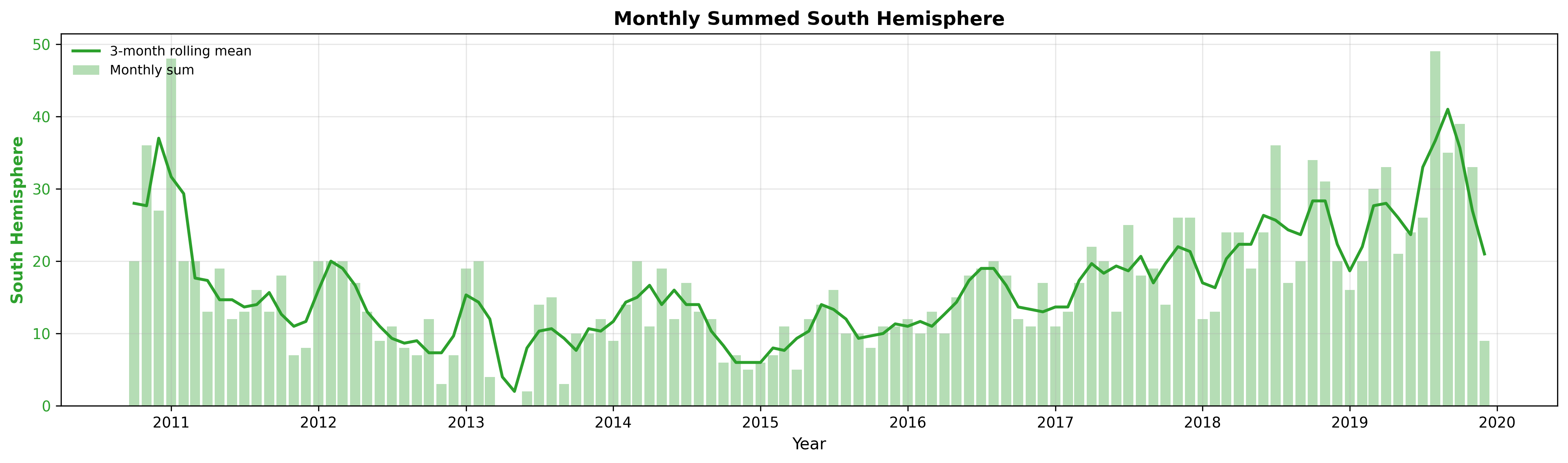}
		\caption{Monthly summed Southern hemispheric newly emerging CH count (bars) \& three-month rolling mean (green line), 2010-2019. The Southern hemisphere is characteristically enhanced in 2010-2011, suppressed through 2013-2015 (near the southern sunspot maximum), and shows a clear long-term increasing trend from 2016 through 2019, with monthly sums reaching their global maximum ($\sim$49) in late 2019 near the Cycle 24/25 minimum.}
		\label{fig:south_monthly}
	\end{figure}

\begin{figure}[H]
		\centering
		\includegraphics[width=0.97\textwidth]{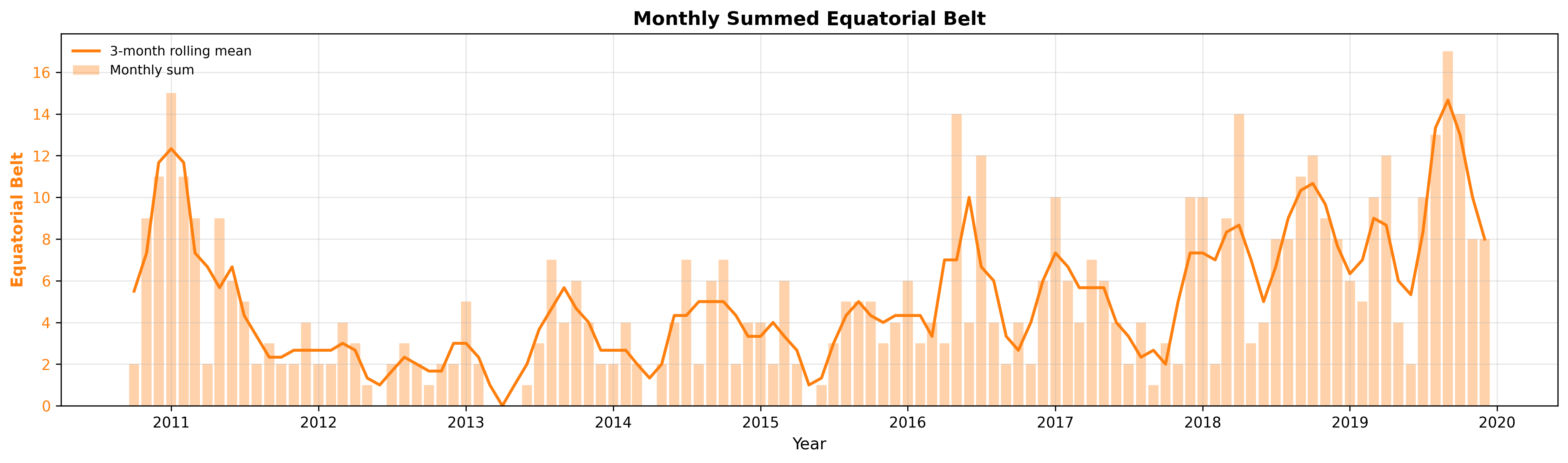}
		\caption{Monthly summed equatorial zone newly emerging CH count (bars) and three-month rolling mean (orange line). The equatorial zone count is maximum in 2010-2011 (monthly sum goes to $\sim$15) and in the late declining phase (2018-2019, reaching $\sim$17), with a prolonged suppression over 2012-2018 broadly coincident with the solar maximum and early decline; this pattern indicates the anti-correlation with the SSN established quantitatively in Section~\ref{sec:eq_ssn}.}
		\label{fig:equator_monthly}
	\end{figure}

	Before showing the derived analyses, we first describe the basic morphology of the newly emerging CH catalogue through monthly summed counts for each latitudinal zone, which we defined earlier in Section~\ref{sec:data:ch}. Figures~\ref{fig:total_monthly}-\ref{fig:equator_monthly} reveal the total, Northern, Southern, and Equatorial zones monthly summed CH counts (bars) and the three-month rolling mean (solid line) spanning the full catalogue. The monthly sum shows the total number of newly emerging CHs reported within each calendar month, whereas the three-month rolling mean is used to emphasize the prolonged trend by smoothing short-term variability. Over all the latitudinal zones, the catalogue shows a distinct temporal profiles that indicates the variation of coronal open magnetic flux by the solar cycle: The total newly emergence count reveal a broad suppression near the solar maxima (2013-2014) and further an rising trend through the declining phase (2016-2019), whereas the northern and southern hemispheric zones demonstrate notably different evolution in temporal profiles, suggesting the hemispheric asymmetry analysis performed in the Section~\ref{sec:asymmetry}. Furthermore, the equatorial zone persistently reveals lower absolute counts than either hemispheric zone over the cycle, demonstrating the fact that newly emerging rate of CH at lower latitudes is less frequent than that at mid and higher latitudes.
	
	\subsection{Equatorial zone newly emerging CH versus sunspot number}
	\label{sec:eq_ssn}
	
\begin{figure}[H]
		\centering
		\includegraphics[width=0.92\textwidth]{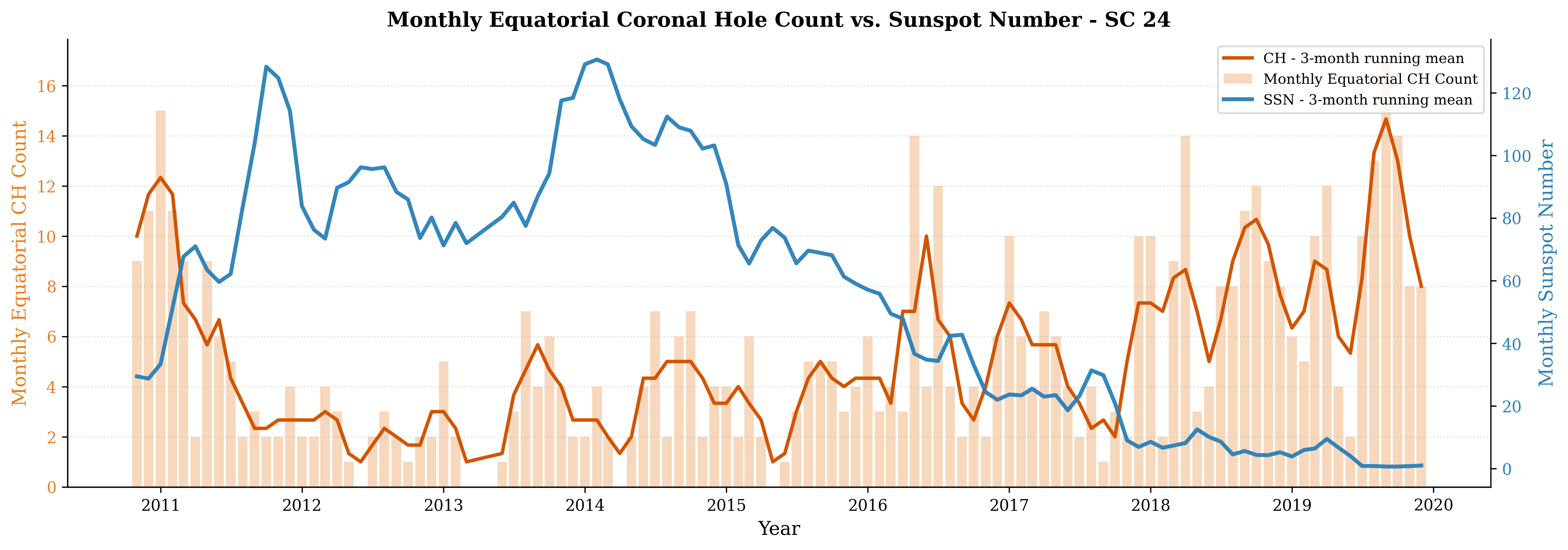}
		\caption{Monthly equatorial CH count, Here x-axis shows the time in years of over Cycle~24, while the left y-axis shows the monthly equatorial coronal hole (CH) count and right y-axis shows the monthly sunspot number (SSN) (orange bars; three-month running mean, orange line, left y-axis) compared with the three-month running mean of monthly SSN (blue line, y-right axis).}
		\label{fig:eq_vs_ssn_trend}
	\end{figure}

	Figure~\ref{fig:eq_vs_ssn_trend} exhibits the three-month running mean of monthly equatorial CH count alongside the three-month running mean of monthly sunspot number throughout the catalogue period. Both time series are visually anti-correlated: equatorial CH emergence is reduced during the years bracketing solar maximum (2013-2015) and enhances during the decreasing phase while approaching solar minimum (2017-2019), reflecting the well-known minimum phase resurgence of equatorial open flux as polar fields restructure \citep{lowder2017,karna2020}.

\begin{figure}[H]
		\centering
		\includegraphics[width=0.62\textwidth]{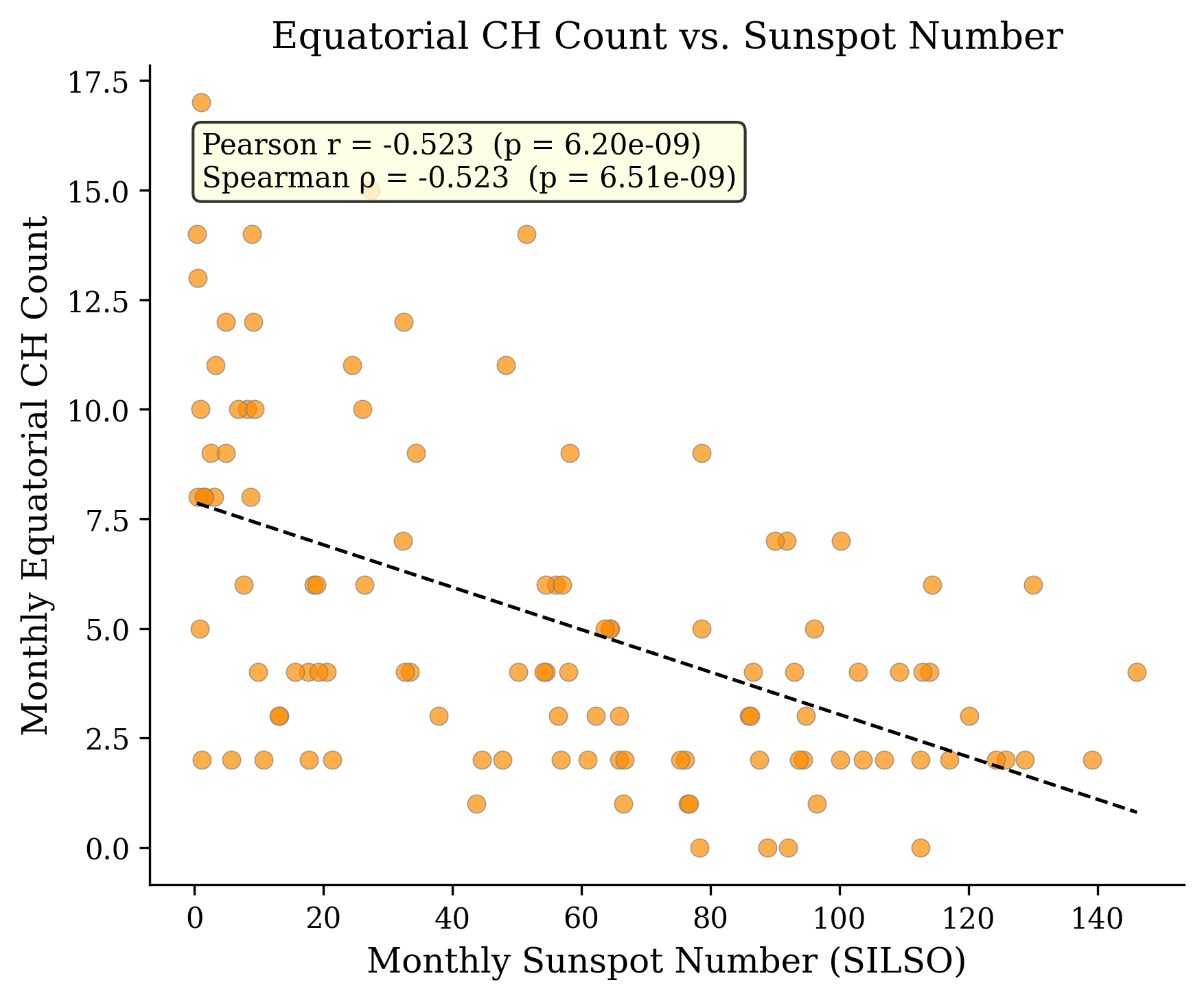}
		\caption{Monthly equatorial CH count versus monthly sunspot number.The x-axis shows monthly sunspot number (SILSO), whereas the y-axis shows monthly equatorial CH count. Dashed line: ordinary least-squares fit.}
		\label{fig:eq_vs_ssn_scatter}
	\end{figure}

	The anti-correlation is statistically significant at monthly intervals: a Pearson correlation of the monthly equatorial CH count benchmarking against the monthly sunspot across $N=108$ overlapping months reveals 	$r = -0.523$ ($p = 6.2\times10^{-9}$), with primarily identical Spearman correlation ($\rho = -0.523$, $p=6.5\times10^{-9}$; Figure~\ref{fig:eq_vs_ssn_scatter}). The best-fit linear relation is $\mathrm{CH}_{\rm eq} = -0.0484\,\mathrm{SSN} + 7.88$, i.e. \ for every $\sim$20 units of monthly SSN increase, the mean number of newly emerging equatorial zone CHs per month reduces by approximately one.

	\subsection{North-South asymmetry of newly emerging coronal holes}
	\label{sec:asymmetry}
	
	\subsubsection{Coronal hole asymmetry index (CH-AI)}

\begin{figure}[H]
		\centering
		\includegraphics[width=0.92\textwidth]{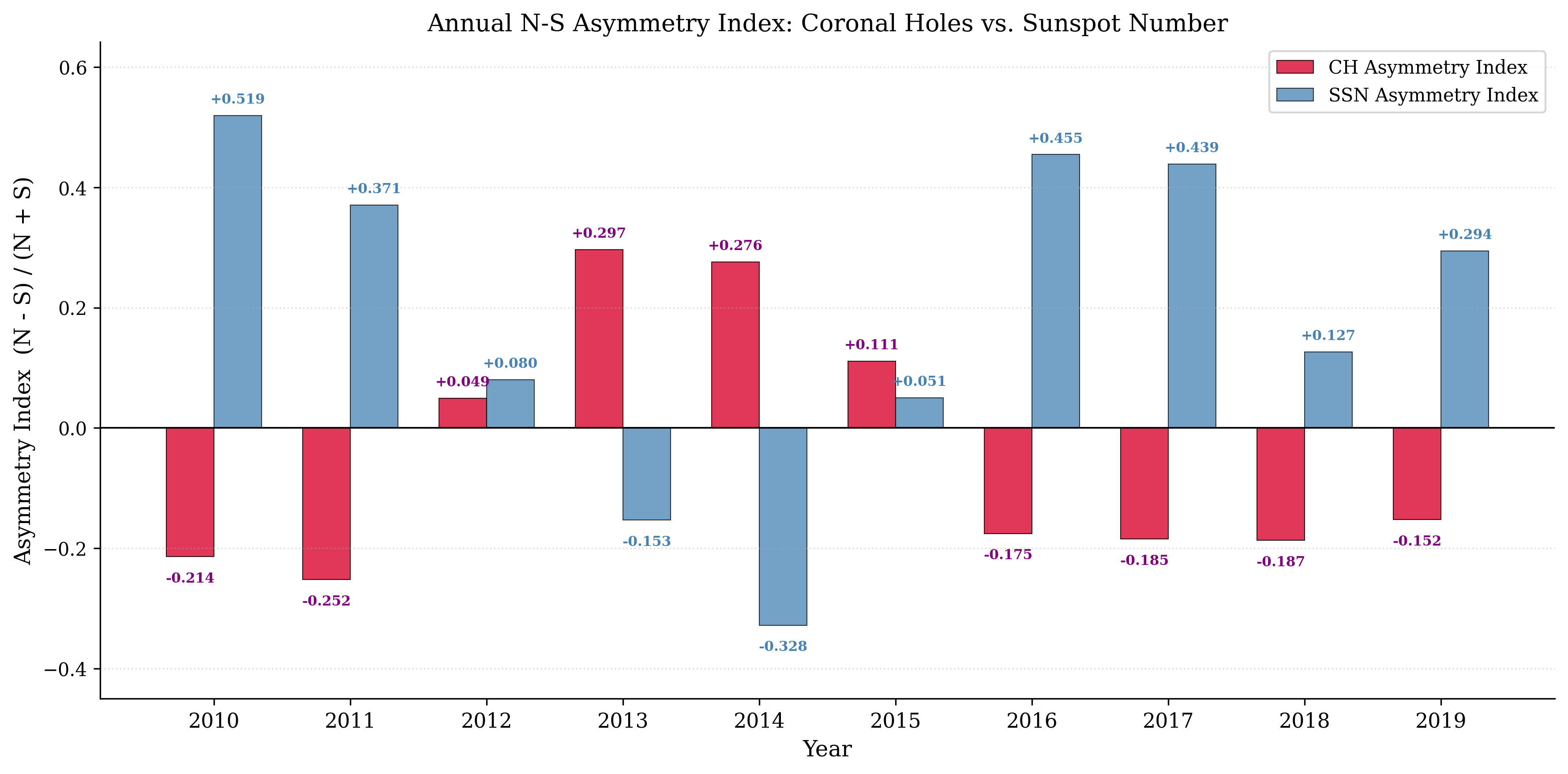}
		\caption{Monthly N-S asymmetry index for CH emergence. The x-axis shows time in years and y-axis shows the asymmetry index (bars; red = north-dominant, green = south-dominant) with a 6-month running mean (dark line) overlaid.}
		\label{fig:ai_timeseries}
	\end{figure}

	The monthly Coronal hole asymmetry index, $\mathrm{AI}_{\rm CH} = (N-S)/(N+S)$, is shown in Figure~\ref{fig:ai_timeseries} collectively with the shading to indicate Northern-dominant (red) and Southern-dominant (green) months. Throughout the full interval, $\mathrm{AI}_{\rm CH}$ has mean $-0.035$ \& standard deviation of $0.298$, with $39$ ($35.8\%$) north-dominant and $68$ months ($62.4\%$) south-dominant.

\begin{figure}[H]
		\centering    
		\includegraphics[width=0.92\textwidth]{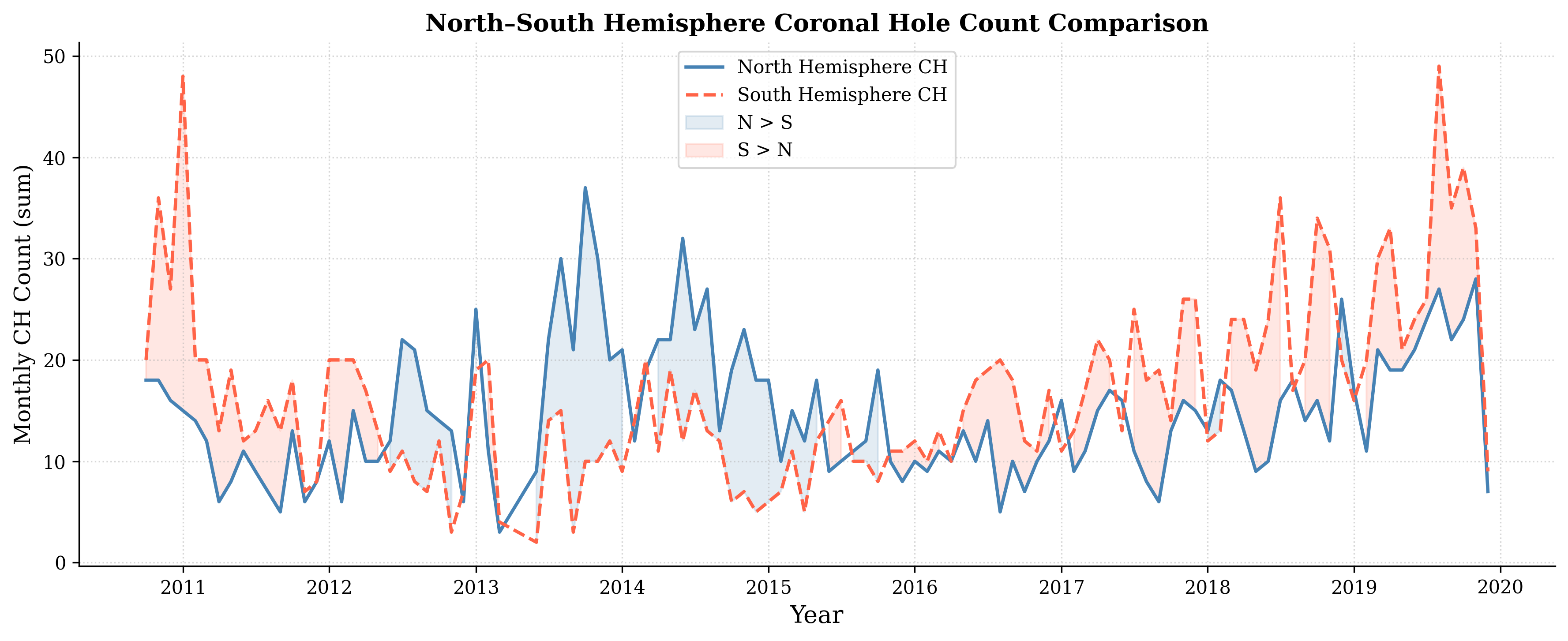}
		\caption{Monthly Northern (blue) and Southern (red dashed) hemisphere CH counts, with shading showing which hemisphere dominates in each month. Here x-axis is showing time in years over Cycle~24, whereas y-axis is showing the monthly CH count (sum). }
		\label{fig:ns_comparison}
	\end{figure}

	However, the sign of $\mathrm{AI}_{\rm CH}$ is far from static: Figure~\ref{fig:ns_comparison} displays that the monthly hemispheric Coronal Hole counts cross repeatedly, with the northern hemisphere dominance for extended period in 2013-2015 (concurrent with the later, enhanced peak of Cycle~24's double maximum) and the southern hemisphere dominance in the initial rising phase (2010-2012) and again through most of the declining phase (2016-2019). The year-by-year breakdown (Table~\ref{tab:annual_dominance}) makes this reversal explicit.
	
	\begin{table}[H]
		\centering
		\caption{Year-by-year monthly-mean hemispheric CH count and dominant hemisphere.}
		\label{tab:annual_dominance}
		\begin{tabular}{@{}lccc@{}}
			\toprule
			Year & Mean North & Mean South & Dominant \\
			\midrule
			2010 & 17.33 & 27.67 & South \\
			2011 & 9.50  & 17.25 & South \\
			2012 & 13.00 & 12.25 & North \\
			2013 & 20.80 & 10.90 & North \\
			2014 & 20.92 & 12.08 & North \\
			2015 & 12.67 & 10.08 & North \\
			2016 & 10.08 & 14.58 & South \\
			2017 & 12.75 & 18.67 & South \\
			2018 & 15.17 & 22.83 & South \\
			2019 & 20.00 & 27.92 & South \\
			\bottomrule
		\end{tabular}
		\\[4pt]

	\end{table}
	
	\subsubsection{Comparison with the sunspot-number asymmetry}
	
	One of the primary questions of this paper is whether the CH asymmetry tracks the well-reported hemispheric asymmetry of SSN. Across Solar Cycle~24, the SSN asymmetry index has a mean of $+0.157$ and a standard deviation of $0.486$ - notably higher in amplitude than the $\mathrm{AI}_{\rm CH}$, and on average north-dominant (opposite in sign to the mean of  CH asymmetry)

\begin{figure}[H]
		\centering
		\includegraphics[width=0.92\textwidth]{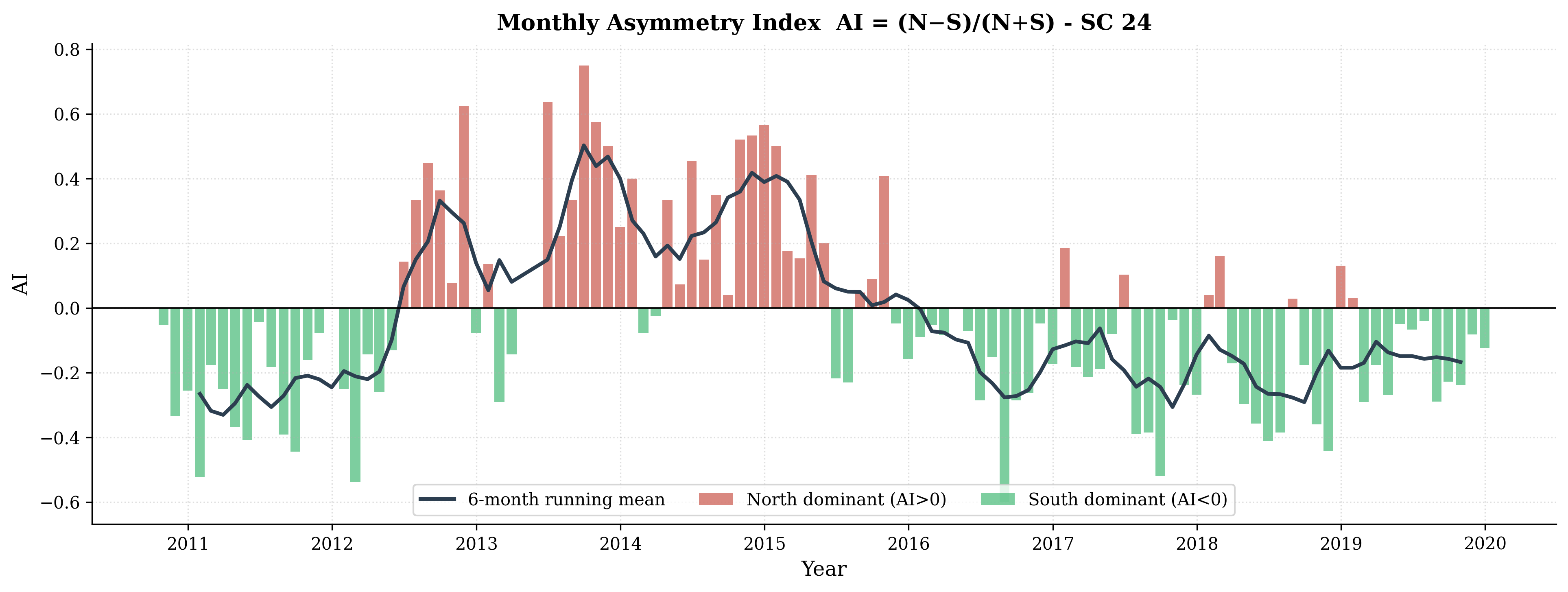}
		\caption{6-month running mean of the CH asymmetry index (red solid) and SSN asymmetry index (blue dashed). The x-axis shows time in year throughout Cycle~24, while the y-axis shows the asymmetry index.}
		\label{fig:ai_comparison}
	\end{figure}

\begin{figure}[H]
		\centering
		\includegraphics[width=0.62\textwidth]{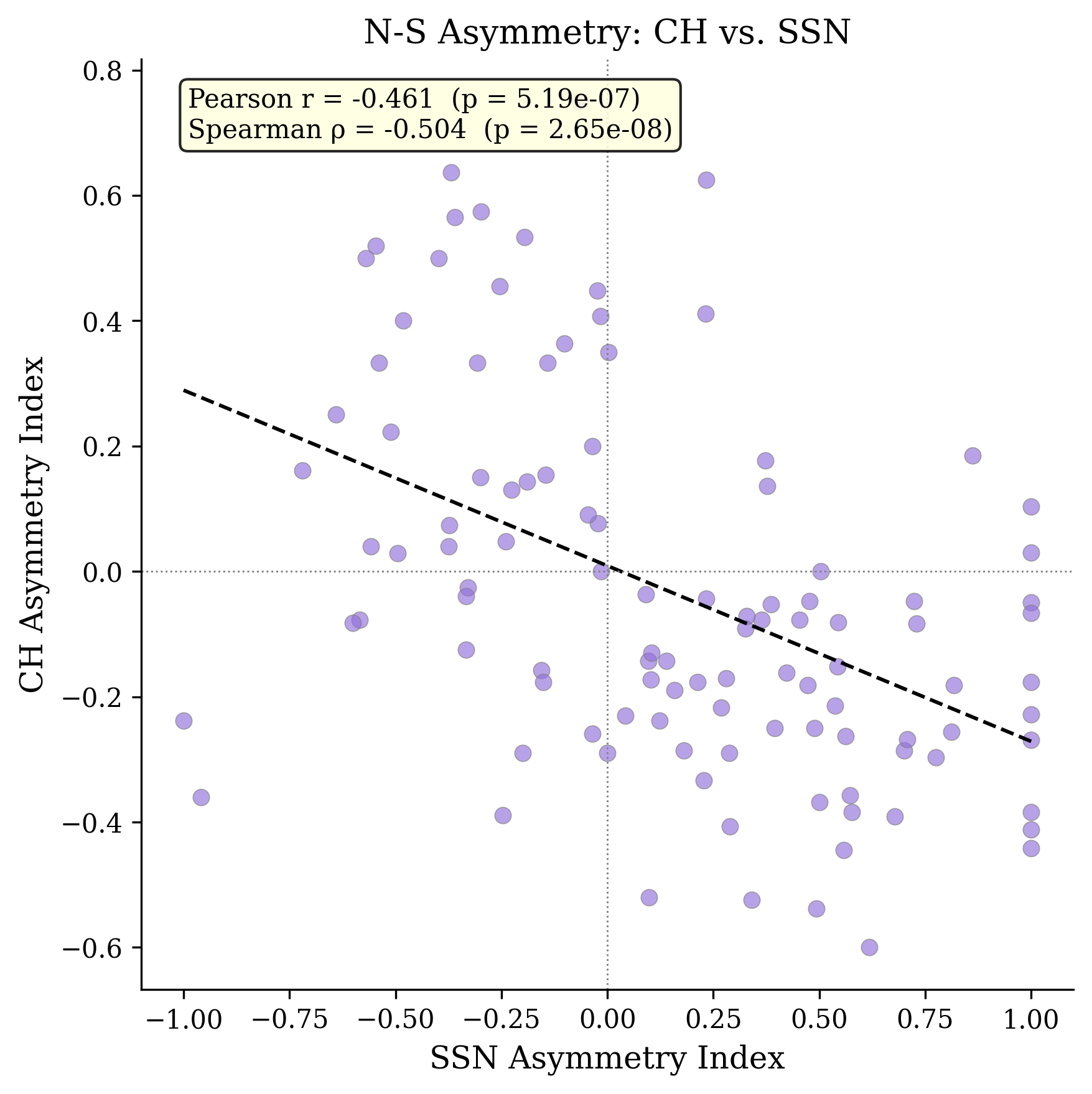}
		\caption{Monthly CH asymmetry index (y-axis) versus monthly SSN asymmetry index (x-axis). Dashed line: ordinary least-squares fit.}
		\label{fig:ai_scatter}
	\end{figure}

	Figure~\ref{fig:ai_comparison} overlays the 6-month running means of the
	$\mathrm{AI}_{\rm CH}$ \& $\mathrm{AI}_{\rm SSN}$. For much of the interval, two indices were visibly out of phase: SSN asymmetry swings strongly positive (northern dominance) in the same interval; the SSN asymmetry then changes sharply towards negative in 2013-2014 (the well recognized late, Southern hemispheric dominated second peak enhancement of the Solar Cycle~24 double maximum) just as the CH asymmetry changes and remains positive (North dominant) through 2015. This anti-phase relationship is driven out quantitatively: the monthly $\mathrm{AI}_{\rm CH}$ and $\mathrm{AI}_{\rm SSN}$ series are markedly anti-correlated (Pearson $r=-0.461$,
	$p=5.2\times10^{-7}$; Spearman $\rho=-0.504$,
	$p=2.7\times10^{-8}$; Figure~\ref{fig:ai_scatter}), with best-fit relation of $\mathrm{AI}_{\rm CH} = -0.280\,\mathrm{AI}_{\rm SSN} + 0.009$.

\begin{figure}[H]
		\centering
		\includegraphics[width=0.92\textwidth]{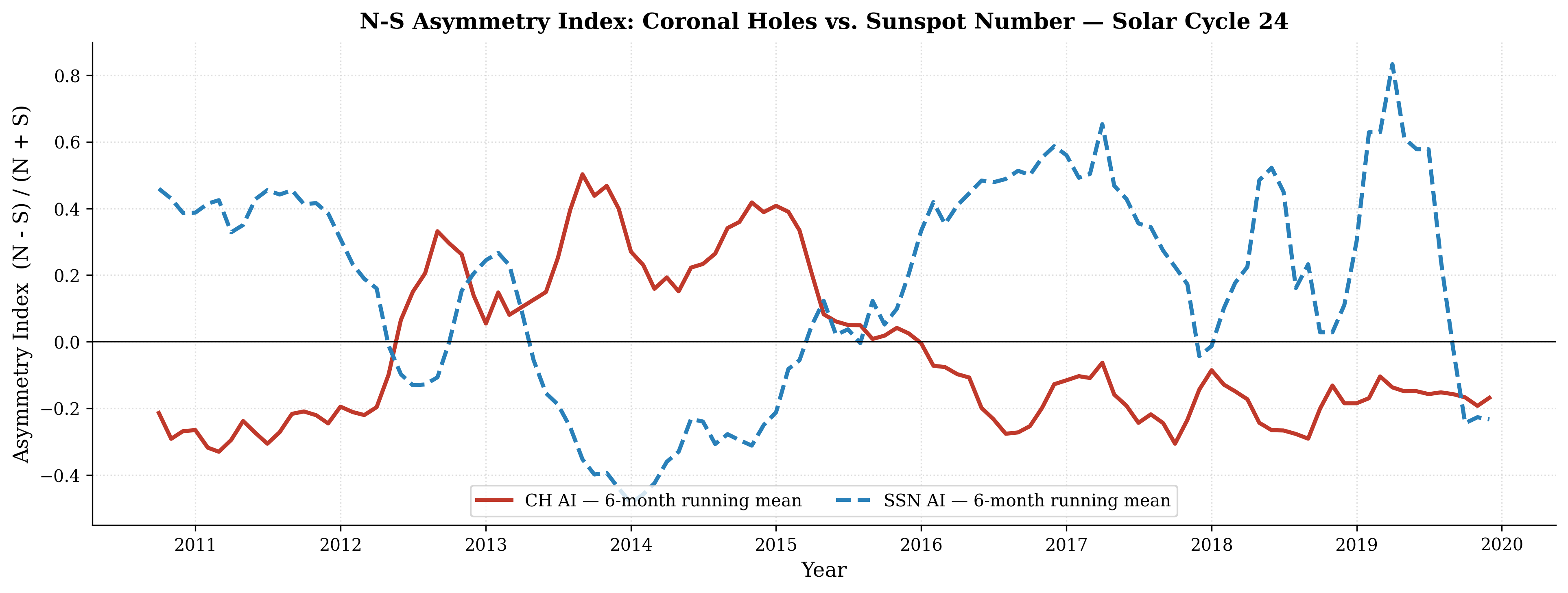}
		\caption{Annual mean CHs asymmetry index (red) \& SSN asymmetry index (blue) by year.}
		\label{fig:ai_annual_bar}
	\end{figure}

	The annual-mean comparison (Figure~\ref{fig:ai_annual_bar}) makes the anti-phase association specifically clear for interval 2013-2014 \&
	2016-2017: in years when the SSN highly favours one hemisphere, the rate of newly emerging CHs tends to favour the other hemisphere. This insight is consistent with the conclusion of \citet{mcintosh2014database} that the Coronal Hole area asymmetry is nearly uncorrelated with sunspot area asymmetry. Our finding enhances this by showing that, over Cycle 24 monthly cadence, specifically two are not merely uncorrelated but significantly anti-correlated.

	\subsection{Solar cycle phase dependence of newly emerging CHs}
	\label{sec:phase}

\begin{figure}[H]
		\centering
		\includegraphics[width=0.92\textwidth]{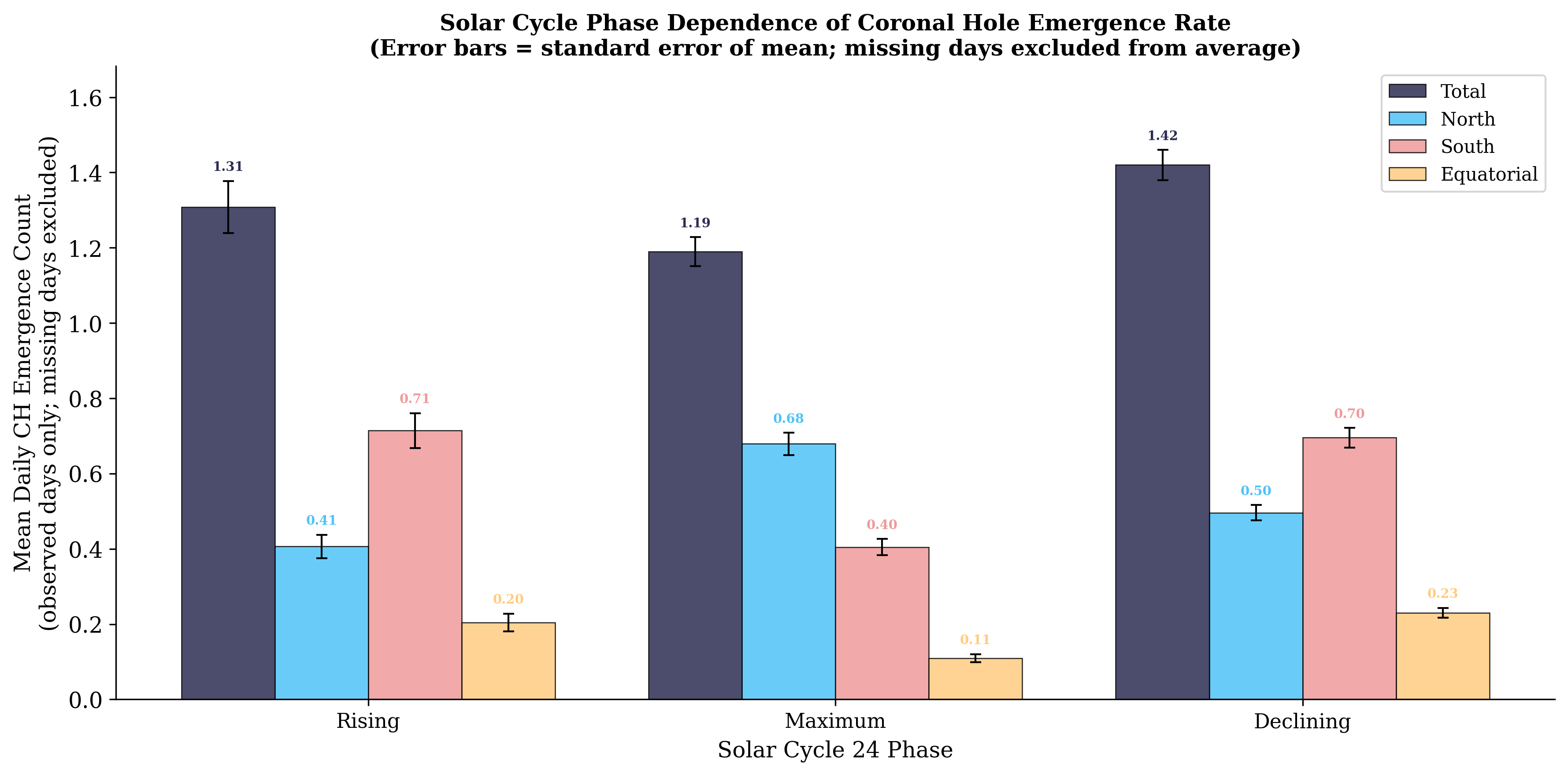}
		\caption{Mean daily CH emergence rate by region (Total, North, South, Equatorial) for each phase of Solar Cycle 24. Error bars = standard error of the mean.}
		\label{fig:phase_bar}
	\end{figure}

	Table~\ref{tab:phase_stats} and Figure~\ref{fig:phase_bar} summarizes
	the mean of daily CH emergence rate, broken down by region, for each of three solar-cycle phases defined in Section~\ref{sec:methods:phase}.
	
	\begin{table}[H]
		\centering
		\caption{Mean daily CH emergence rate by solar-cycle phase and region (errors = standard error of mean).}
		\label{tab:phase_stats}
		\begin{tabular}{@{}lcccccc@{}}
			\toprule
			Phase & $N_{\rm obs}$ & $N_{\rm all}$ & Missing & Total & North & South \\
			\midrule
			Rising     & 496  & 557  & 11.0\% & $1.309 \pm 0.069$ & $0.406 \pm 0.031$ & $0.714 \pm 0.046$ \\
			Maximum    & 981  & 1204 & 18.5\% & $1.190 \pm 0.039$ & $0.679 \pm 0.029$ & $0.405 \pm 0.022$ \\
			Declining  & 1546 & 1645 &  6.0\% & $1.420 \pm 0.040$ & $0.496 \pm 0.020$ & $0.696 \pm 0.026$ \\
			\bottomrule
		\end{tabular}
		\\[6pt]
		\begin{tabular}{@{}lc@{}}
			\toprule
			Phase & Equatorial \\
			\midrule
			Rising     & $0.204 \pm 0.024$ \\
			Maximum    & $0.109 \pm 0.011$ \\
			Declining  & $0.230 \pm 0.013$ \\
			\bottomrule
		\end{tabular}
		\\[4pt]
		\footnotesize{Note: $N_{\rm obs}$ = days with a recorded observation; $N_{\rm all}$ = calendar days in the phase interval.}
	\end{table}
	
	Two characteristics of Table~\ref{tab:phase_stats} stand out. The first one is that the total rate of newly emerging CHs is lowest throughout the maximum phase ($1.190$ CH day$^{-1}$, observed days only) and highest throughout the declining phase ($1.420$ CH day$^{-1}$), consistent with the long-standing picture that hemispheric polar open flux - and thus the supply of newly emerging CHs is replenished as the cycle falls towards the next minimum phase \citep{karna2020,lowder2017}.

	Second one is the equatorial zone CH emergence rate is itself phase-dependent in a direction stable with the anti-correlation with the SSN validated in Section~\ref{sec:eq_ssn}: lowest at the solar maximum interval ($0.109$ day$^{-1}$) and highest throughout the declining phase ($0.230$ day$^{-1}$), with rising phase intermediate ($0.204$ day$^{-1}$).

\section{Discussion}
\label{sec:discussion}

\subsection{Equatorial CH emergence as an anti-correlated tracer of solar activity}

The significant anti-correlation we identified between equatorial newly emerging CHs and the SSN ($r \approx -0.52$ at monthly cadence) reinforces, with an independent and methodologically unique data record, the widely accepted view that the lower-latitude open magnetic flux is suppressed when there are more active region (closed flux) present on the photosphere, and is released as those active region further decays \citep{karna2020,lowder2017,bravo1997}.
The catalogue that we reported was based on the counts of newly emerging CHs rather than the number or area of the CHs present on a given day over the solar disk, as it is more insensitive to the lifespan of individual CHs and instead probes the rate of new lower-latitude open magnetic flux formation. Hence, this gives a novelty to our results based on the unique methodology we used rather than qualitative analysis.

\subsection{The CH asymmetry is not simply inherited from sunspots}

Perhaps our major substantial result is that the CH hemispheric asymmetry is significantly anti-correlated with the SSN hemispheric asymmetry (Section~\ref{sec:asymmetry}),  rather than being uncorrelated (as reported for CH area in McIntosh Archive analysis of \citealp{mcintosh2014database}). 
The probable contributor to this phase opposed relationship is the well-archived $\sim$2-year lag between the northern and southern hemispheric sunspot maxima throughout the Solar Cycle~24: the SSN (SILSO) hemispheric study reports that first (northern) peak of the Cycle~24 double maxima happened around November 2011 and the second, enhanced (southern) peak happened in the early 2014, with the lagging in southern hemisphere by nearly $26$ months \citep{silso2024}.
If the open magnetic flux around polar/hemispheric region is replenished preferably in the hemisphere that has priorly passed its local sunspot maximum and started returning to a dipole predominated state, the hemisphere with prior (northern) sunspot peak would be expect to develop an increase in CH count before the hemisphere with the later (southern) peak does – which is consistent with the behaviour observed in Figure~\ref{fig:ai_comparison}, where the CH asymmetry swings positive (northern-dominated) in 2013-2015, following the northern sunspot peak but bracketing the southern one

\subsection{Comparison with prior coronal hole asymmetry studies}

The results we have shown of an overall southern bias in emergence in CH ($62\%$ of months Southern-dominated versus $36\%$ northern-dominated) are consistent with the CH area-based outcomes reported by \citet{nakagawa2019}, who observed a north-south (N-S) asymmetry in the area of CH, where the area of CH was larger in the southern hemisphere than the northern hemisphere throughout the solar cycle 23 and 24
It is also consistent with the asymmetry in the area of polar CH for Solar Cycle~24 by \citet{andreeva2021}, who reported a polar-CH-area asymmetry in both northern and southern hemispheres that needs further explanation, whereas the area of non-polar CH fluctuates quasi-synchronously with sunspot activity. As our results differ from these reported ones, however, our North/South categories merge polar and non-polar CHs also (Section~\ref{sec:data:ch}).

\section{Conclusion}
\label{sec:conclusion}

We reported a novel, long-term, manually validated catalogue of newly emerging CHs daily and then further classified them based on hemisphere and latitudinal zone for Solar Cycle~24, derived from the CHIMERA/SolarMonitor maps. Using this catalogue, we revealed that

\begin{enumerate}
	\item The emergence of newly equatorial CH is substantially anti-correlated with the SSN at monthly cadence ($r=-0.523$, $p=6.2\times10^{-9}$), validating with an independent, emergence-rate-based dataset the verified area-based picture that lower-latitude open magnetic flux is suppressed near the solar maxima. 
	\item We also observed the North-South (N-S) asymmetry of newly emerging CHs throughout the cycle, weighted toward the southern hemisphere ($62\%$ of months south-dominant), also consistent with previous area-based research studies of Solar Cycle~24, but the asymmetry reverses sign for a sustained interval (2013-2015) concurrent with the second, stronger peak of Solar Cycle~24’s double maximum.
	\item The asymmetry index of CH is considerably anti-correlated with the asymmetry index of SSN ($r=-0.461$, $p=5.2\times10^{-7}$), suggesting that emergence of CH does not simply mirror the hemisphere of higher sunspot activity – if anything, the opposite hemisphere is inclined to show increased CH emergence, a trend plausibly related to the well-documented $\sim$26-month lag between the northern and southern sunspot maximum in Solar Cycle~24 
	\item The total CH emergence is majorly modulated by cycle phase, minimum at solar maximum ($1.19$ CH day$^{-1}$, observed days only) and highest over the declining phase ($1.42$ CH day$^{-1}$, while the equatorial portion subsequently follows the same phase dependence as the anti-correlation with SSN – reduced at maximum ($0.11$ day$^{-1}$) and identical between the rising and declining phases ($0.20$ vs $0.23$ day$^{-1}$; $p=0.13$).
\end{enumerate}
 These results show that a manually compiled catalogue of newly emerging CH is distinct from the existing area-based catalogues and literature. This helps to understand the features of CH-cycle and CH-hemisphere relationships, which also reveals the unreported anti-phase relationship between the CH and SSN hemispheric asymmetries throughout the Solar Cycle~24.

\section*{Acknowledgements}

The authors thank the CHIMERA team and the SolarMonitor team (Dublin Institute for Advanced Studies, Trinity College Dublin, and Northumbria University) for making the coronal hole daily segmentation data publicly available at \url{https://www.solarmonitor.org}. The authors also express gratitude to the SILSO World Data Center, Royal Observatory of Belgium (ROB), Brussels, for providing the international sunspot number data.

\section*{Data Availability}

The Newly Emerging Coronal Hole data catalogue compiled in the research study is publicly available as supplementary material on Zenodo at \url{https://doi.org/10.5281/zenodo.21968544}. Other data, including sunspot number data of total count and hemispheric count (SILSO) and coronal hole maps data (CHIMERA) are also publicly available on their official archives.

\bibliographystyle{plainnat}
\bibliography{reference}

\end{document}